\documentclass[sigconf]{acmart}

\AtBeginDocument{%
  \providecommand\BibTeX{{%
    \normalfont B\kern-0.5em{\scshape i\kern-0.25em b}\kern-0.8em\TeX}}}

\usepackage{multirow}

\begin{document}

\title{Actor-based Risk Analysis for Blockchains in Smart Mobility}
\subtitle{Forthcoming in the proceedings of CryBlock/MobiCom2020}


\author{Ranwa Al Mallah}
\affiliation{%
  \institution{Ryerson University}
  \streetaddress{350 Victoria St, P.O. Box M5B 2K3}
  \city{Toronto}
  \state{Ontario}}
\email{ranwa.almallah@ryerson.ca}

\author{Bilal Farooq}
\affiliation{%
  \institution{Ryerson University}
  \streetaddress{350 Victoria St, P.O. Box M5B 2K3}
  \city{Toronto}
  \state{Ontario}}
\email{bilal.farooq@ryerson.ca}


\begin{abstract}
  Blockchain technology is a crypto-based secure ledger for data storage and transfer through decentralized, trustless peer-to-peer systems. Despite its advantages, previous studies have shown that the technology is not completely secure against cyber attacks. Thus, it is crucial to perform domain specific risk analysis to measure how viable the attacks are on the system, their impact and consequently the risk exposure. Specifically, in this paper, we carry out an analysis in terms of quantifying the risk associated to an operational multi-layered Blockchain framework for Smart Mobility Data-markets (BSMD). We conduct an actor-based analysis to determine the impact of the attacks. The analysis identified five attack goals and five types of attackers that violate the security of the blockchain system. In the case study of the public-permissioned BSMD, we highlight the highest risk factors according to their impact on the victims in terms of monetary, privacy, integrity and trust. Four attack goals represent a risk in terms of economic losses and one attack goal contains many threats that represent a risk that is either unacceptable or undesirable. 

\end{abstract}



\maketitle

\section{Introduction}
Nowadays, transportation data are shared across multiple entities and stored in central servers that are susceptible to cyberattacks. In the recent years, many cybersecurity breaches have occurred in transportation systems. In 2015, a group of civic hackers deciphered and exposed the unstandardized bus system location data of Baltimore \cite{rector2018mta}. The San Francisco transit was hacked to give free access to commuters for two days \cite{stewart2016sf}. It was recently discovered that Google keeps collecting user location data even if users explicitly deactivate the tracking system in their mobiles \cite{nakashima2018ap}.  

Blockchain technology creates a distributed consensus, providing entities with a secure platform that maintains past records of digital events \cite{hasanova2019survey}. In the context of transportation, a multi-layered Blockchain framework for Smart Mobility Data-market (BSMD) was recently proposed by L\'opez and Farooq  \cite{lopez2020multi}. BSMD is a permissioned blockchain and is designed to solve the privacy, security and management issues related to the sharing of passively as well as actively solicited large-scale mobility data. In another study, L\'opez and Farooq \cite{9071730} proposed a distributed tool for mobility choice modelling over BSMD, where participants do not share personal raw data, while all computations are done locally. Eckert et al. developed a user-centric emission monitoring and trading system for multi-modal mobility over the BSMD \cite{eckert2019blockchain}. Another application in transportation used BSMD for mode choice inference using federated learning over blockchain \cite{Lopezetal2019}.

However, security issues related to blockchain are critical in terms of cybersecurity. In this sense, security experts need to fully understand the risk in terms of scope and impact of the security and privacy challenges related to blockchain before predicting the potential damage from an attack. Furthermore, verifying whether the current technology can withstand persistent hacking attempts is also of utmost importance.

In 2017, Li et al. conducted a comprehensive cybersecurity risk analysis of blockchain \cite{li2017survey}. They systematically studied the security threats to blockchain and survey the corresponding real attacks by examining popular blockchain systems. Although they identified the threats and their nature, the real scope of the risk that those threats entail was not described and evaluated. Unlike their work, we aim at quantifying the risk. We present ordinal values for the identified risks. Another difference is that in our work, we analyze a real blockchain system developed for transportation application i.e. Blockchain for Smart Mobility Data-markets. 

The aim of our work is more extensive than the traditional risk assessment. We estimate the risk through an actor-based risk analysis based on the impact that the exploitation of vulnerabilities will have on the transportation systems using BSMD. Efficient detection mechanisms are desired in this context to reduce the risk that various threats entail on the blockchain systems used in transportation domain. Countermeasures to avoid cyberattacks must be implemented as a security-by-design practice. 

This paper is organized as follows. Related work is provided in Section 2. In Section 3, we present the methodology followed by the actor-based risk analysis of a blockchain in transportation domain in section 4. Finally, conclusion and future work are outlined in Section 5.

\section{Related work}

In the last decade, several studies have exposed vulnerabilities in the technologies employed in blockchain. Particularly, for research in blockchain cybersecurity risk analysis, we summarize the findings in Table \ref{relatedwork}.

\begin{table*}[h]
	\centering
	\caption{Previous studies in blockchain cybersecurity risk analysis.}
	\renewcommand{\arraystretch}{0.5}
	\begin{tabular}{|p{0.25\linewidth}|p{0.34\linewidth}|p{0.3\linewidth}|}
    \hline
\textbf{Studies} & \textbf{Methodology} & \textbf{Drawbacks}\\
    \hline
    \multirow{3}{*}{Li et al.,  \cite{li2017survey}} & - Systematic examination on security risks for popular blockchain systems. & - The real scope of the risk that those threats entail is not described. \\
    & - Survey the real attacks on blockchain systems (Ethereum, Bitcoin, Monero, RSK, Counterpaty, Stellar, Monax, Lisk). & - Do not consider the risk as being a function of probability and impact.\\
& - Analyze related vulnerabilities exploited. & \\
    \hline
    \multirow{3}{*}{Atzei et al., \cite{atzei2017survey}} & - Analyzed the security vulnerabilities of Ethereum smart contracts.
 & - Isolated the analysis from a security programming perspective only.
 \\
& - Show a series of attacks which exploit these vulnerabilities.
 & - They do not account for the various nature of security threats.
 \\
    \hline
    \multirow{3}{*}{Homoliak et al., \cite{homoliak2019security}} & - Stacked hierarchy of four layers.
 & - They don’t quantify the risk.\\
& - Identified four threat agents.
 & - Generic architecture was not evaluated on a realistic blockchain system. \\
 & - Report vulnerabilities at each layer. &  \\
    \hline
\end{tabular}
	\label{relatedwork}
\end{table*}

Li et al. \cite{li2017survey} performed a systematic examination of the risks associated with the popular blockchain systems (i.e. Ethereum, Bitcoin, Monero, RSK, Counterpaty, Stellar, Monax, Lisk), the corresponding real attacks, and the security enhancements. They evaluated the real attacks on popular blockchain systems from 2009 to 2017 and analyzed the vulnerabilities exploited in these cases.

Rather than popular blockchain systems, Atzei et al. \cite{atzei2017survey} focused on Ethereum smart contracts. From a security programming perspective, their work analyzed the security vulnerabilities of Ethereum smart contracts, and provided a taxonomy of common programming pitfalls that may lead to vulnerabilities. They show a series of attacks on smart contracts that exploit these vulnerabilities, allowing an adversary to steal money or cause other damage. 

In contrast to previous studies, Homoliak et al. \cite{homoliak2019security} proposed a security reference architecture based on models that demonstrate the stacked hierarchy of various threats (similar to the ISO/OSI hierarchy). In order to isolate the various nature of security threats, the stacked hierarchy consisted of four layers (Network layer, consensus layer, replicated state machine layer and application layer). They also proposed a threat-risk assessment of the reference architecture using the ISO/IEC 15408 template. With the stacked model, different threat-agents and threats appear at each layer. They identified four threat-agents: service providers, consensus nodes, developers and users. They are malicious entities whose intention is to steal assets, break functionalities, or disrupt services. 

Although the previous studies identified the threats and their nature, the real scope of the risk that those threats entail is not described. We consider that although there are vulnerabilities in a system, it is the impact that certain exploitation has on the nodes of the blockchain network that determines whether the vulnerability represents a significant risk or not. In this line of thought, we have carried out an actor-based risk analysis that we applied to the multi-layered blockchain for smart mobility data-market. 

\section{Methodology}

The aim of our work is to perform an actor-based risk analysis. It is more extensive than the traditional risk assessment in the manner that the methodology enables us to quantify the risk. Quantifying the risk refines the analysis and gives an order of magnitude of the risk exposure. Also, measuring the risk enables to better grasp the impact of the attacks on large-scale systems or ecosystems where many technologies are to be evaluated by the risk assessment.

In the following, we define some cybersecurity and risk assessment terms in order to facilitate the reading of our work. 

\begin{itemize}
\item \textbf{Actor:} An entity that violates integrity, privacy or confidentiality to obtain certain benefit.

\item \textbf{Victim:} An entity that is the subject of a cyberattack.

\item \textbf{Attack goal:} Final effect desired by the actor aiming on producing an impact on the victim.

\item\textbf{Scenario:} Set of actions carried out by the actor to achieve its attack goal.

\item \textbf{Impact:} Quantification of the attack goal’s effect on the victim. 

\item \textbf{Threat:} Often called the actor scenario pair, is the combination of the entity who commits the act, the actor, and the way it is committed, the scenario, in order to produce a negative impact.

\item \textbf{Vulnerability:} A flaw that offers the opportunity to damage a system. 

\end{itemize}

The first step of the actor-based risk analysis is to identify potential attackers, i.e. actors who would be interested in the ecosystem under study. The next step is to determine the attack goals of the actors. Finally, Table \ref{impactlevels} is used to quantify the impact on the victim of such attack goals according to a four-level scale.

We follow this methodology to determine the impact that results from attacks on the BSMD ecosystem, a realistic blockchain system. Specifically, Lopez and Farooq \cite{lopez2020multi} proposed the Blockchain for Smart Mobility Data-markets, which is composed of nodes: Individuals, Companies, Universities and Government (transport, census, planing and development agencies). The nodes collect their own data and store it in identifications. Each node is the sole owner of their data and can share their information by showing other nodes their identifications or parts of it. In the blockchain there are smart contracts available that the nodes need to sign before any transaction of information is conducted. An actor-based risk analysis is conducted on this system.

\begin{table*}[t]
         \caption{ Impact levels for mobility data-market adapted from ICS-CERT \cite{GAO}}
	\centering
	\renewcommand{\arraystretch}{0.5}
	\begin{tabular}{|c|c|l|}
    \hline
\textbf{Type} & \textbf{Level} & \textbf{Description of the impact}\\
    \hline
    \multirow{4}{*}{Monetary} & 1 & Minor monetary loss \\
& 2 & Significant monetary loss  \\ 
& 3 & Severe monetary loss \\
& 4 & Catastrophic monetary loss \\
\hline
    \multirow{4}{*}{Privacy} & 1 & Minor impact on the privacy of any of the nodes \\
    && in BSMD (Individuals, Companies, Universities \\
    && and Government) \\
& 2 & Significant on the privacy of the nodes in BSMD  \\ 
& 3 & Severe on the privacy of the nodes in BSMD \\
& 4 & Catastrophic on the privacy of the nodes in BSMD \\
\hline
\multirow{4}{*}{Integrity} & 1 & Minor impact on the integrity of the mobility data, \\
&& transactions and integrity of the users \\
& 2 & Significant impact on the integrity of the mobility \\
&& data, transactions and integrity of the users  \\ 
& 3 & Severe impact on the integrity of the mobility data, \\
&&
transactions and integrity of the users \\
& 4 & Catastrophic impact on the integrity of the mobility \\
&& data, transactions and integrity of the users \\
\hline
    \multirow{4}{*}{Trust} & 1 & Minor impact on the trust of the BSMD network \\
& 2 & Significant  impact on the trust of the BSMD network   \\ 
& 3 & Severe impact on the trust of the BSMD network \\
& 4 & Catastrophic impact on the trust of the BSMD \\
\hline
\end{tabular}
    \label{impactlevels}
\end{table*}

\section{Actor-based risk Analysis of BSMD}

The methodology in our work can be applied to any cybersecurity analysis. Here, we describe its application on BSMD, a cyber physical transportation system, but the risk evaluation methodology is not constrained to blockchain and can be applied to other technologies or generic blockchain systems. 

In the first step, we need to identify cyber threat sources (potential attackers) and their attack goals in the context of smart mobility data-markets.

\subsection{Cyber threat sources}
Cyber attackers exploit the systems for financial gains, obtaining information, conducting sabotage activities, creating disinformation, and degrading confidence in the ecosystem. The Industrial Control Systems Cyber Emergency Response Team  (ICS-CERT) has characterized a cyber threat source as: ``persons who attempt unauthorized access to a control system device and/or network using a data communications pathway \cite{GAO}.'' It further classifies these threat sources into four groups (A1 through A4). The taxonomy of cyber threat sources was introduced for traditional threats to purely Information Technology (IT) infrastructure. We propose to use the same taxonomy in the context of cyber threats against the BSMD ecosystem---since it is composed of both IT elements and of cyber physical systems that have an IT component. In the context of threats against the BSMD ecosystem, we added a new threat group that we call ``insider threat''. We suppose it is a determined node with elevated economic means and high motives. The details of various actors are described below: 

\textbf{A1. Cybercriminals:} This group includes traditional cybercriminals that use compromised computer systems to commit identity theft and leverage the blockchain network for a variety of malicious activities, mostly for monetary gain. They may be passive nodes such as small businesses, individuals, data collectors, or others. 

\textbf{A2. Industrial spies:} Organizations that compromise the computer systems to illegally acquire intellectual property, know-how, trade, and commercial secrets, or other kinds of corporate confidential information. This kind of espionage may occur between competing corporations, for economic reasons. 

\textbf{A3. Foreign Intelligence Agencies:} Foreign state-based organizations that use computer systems to acquire sensitive information on opposing states, corporations or individuals, or otherwise influence their actions. 

\textbf{A4. Terrorist groups:} Organizations seeking to create public disorder or sow national terror, by committing destructive violent acts.

\textbf{A5. Insider threat:} The insider threat is a corrupted active node of the blockchain network (Government, University, Company or other), or an infrastructure node such as a network operator, an Internet Service Provider (ISP) that maliciously exploits the blockchain system.

\subsection{Attack goals}
We identify five attack goals and present them in Table \ref{sumGoals}. We describe them herein.

\begin{table*}[h]
	\centering
	\caption{Attack goals against the blockchain framework for smart mobility data-markets}
	\renewcommand{\arraystretch}{0.5}
	\begin{tabular}{|p{0.3\linewidth}|p{0.5\linewidth}|}
    \hline
\textbf{Goal} & \textbf{Example}\\
    \hline
    G1 - Gain knowledge about the data-market & In 2015, a group of civic hackers deciphered and exposed the unstandardized bus system location data of Baltimore \cite{rector2018mta}. \\
    \hline
    G2 - Access sensitive data on the nodes of the network & It was recently discovered that Google keeps collecting user location data even if they explicitly deactivate the tracking system in their mobiles \cite{nakashima2018ap}. In 2016, information of 57 million Uber customers and drivers were leaked \cite{wong2017uber}.\\
    \hline
    G3 - Manipulate and modify blockchain information & In 2016, criminals manipulated Smart contracts in the Ethereum blockchain with a DAO hack (Decentralized Autonomous Organization), to steal around 60 million dollars \cite{mehar2019understanding}. \\
    \hline
    G4 - Sabotage activities &  In 2016, the San Francisco transit was hacked to give free access to commuters for two days \cite{stewart2016sf}.\\
    \hline
    G5 - Induce participants in the blockchain network to make errors & In the Bitcoin network, with a double spending attack on fast payments, it was shown in \cite{karame2012double} that nodes in the networks may not detect an invalid transactions and add it into the secure ledger. The attacker thus enjoys a service without paying. \\
    \hline
\end{tabular}
	\label{sumGoals}
\end{table*}

\textbf{G1 - Gain knowledge about the data-market}

There is significant competition between companies that produce transportation information. For example, telecommunication companies generate data that can be used for transportation modeling. Similarly, the logs of available mobile devices registered by cellphone towers can be used to monitor traffic. Companies are in control of their data. Accordingly, the BSMD ecosystem could be a target for industrial spies (A2) aiming to obtain knowledge about the data-market. Subsequently, such information could be sold to competing companies. Furthermore, this information is also valuable for criminal groups (A1), intelligence services (A3) and terrorist groups (A4) because it allows them to undertake attacks by maliciously exploiting the knowledge about the data-market.

\textbf{G2 - Access sensitive data on the nodes of the network}

Mobility data is continuously generated by different nodes of the BSMD network. According to the identification layer of the multi-layered blockchain model for smart mobility data market, transportation data is stored in files called Identification. Each entity contains metadata, static data and dynamic data. Nodes need consent from the owner to access the static and dynamic data which can be attractive to many actors. 

On the other hand, nodes communicate with each other using Decentralized Identifiers (DID). DIDs are the gates for sharing data via peer-to-peer connections where the information is transferred using an asymmetric encryption. DIDs represent sensitive information because a given node will have one unique DID per transaction. Any leakage of this information enables to correlate DIDs in the ledger to track single nodes and obtain information on transactions. Intelligence services (A3) and terrorist groups (A4) would be interested in having this information because it would allow them to attain their ultimate goal of surveillance, assassination, etc. Cybercriminal groups (A1) would be interested in this information to obtain monetary gain since mobility data is highly valued. Through unsolicited sharing of information, their clients could be for example insurance companies (medical or automotive) that may use information on an individual’s daily activity patterns to assess the cost of insurance premiums or simply refuse coverage. Industrial spies (A2) may take advantage of the blockchain to find customers and improve their business.

\textbf{G3 – Manipulate and modify blockchain information}

Blockchains are made to be practically immutable, where, in theory no one can modify the blockchain's 'distributed ledger' of all committed blocks. The blockchain relies on the distributed consensus mechanism to achieve immutability and to establish mutual trust. Vulnerabilities in the consensus mechanism can be exploited by attackers to control the blockchain by manipulating and modifying the blockchain information. Criminal groups (A1), intelligence services (A3) and terrorist groups (A4) may want to reverse, exclude or modify the ordering of transactions. An insider threat (A5) may hamper normal mining operations of other miners or impede the confirmation operation of normal transactions. 

\textbf{G4 – Sabotage activities}

Cybercriminals (A1) may use the blockchain ecosystem for sabotage activities. Since the process is anonymous, it is hard to track user behaviors, let alone subject to legal sanctions. For example, criminals can leverage smart contracts for a variety of malicious activities, which may pose a threat to our daily life. Criminal Smart Contracts can facilitate the leakage of confidential information, theft of cryptographic keys, and various real-world crimes. 

Sabotage activities may also be conducted by intelligence services (A3) and terrorist groups (A4) to disrupt the blockchain network or make victims view of the blockchain filtered. As programs running in the blockchain, smart contracts may have security vulnerabilities caused by program defects. Attackers may exploit the vulnerabilities to send malware and infect nodes or to initiate Denial of Service (DoS) attacks on the nodes of the blockchain. The attacks may cause a waste of hard disk resources and decreased node speed. An insider threat (A5) may be a corrupted node who wants to intercept the network traffic of the blockchain to potentially delay network messages. An attacker may want to monopolize all of the victim's incoming and outgoing connections, which isolates the victim from the other peers in the network. Then, the attacker can filter the victim's view of the blockchain, or let the victim cost unnecessary computing power on obsolete views of the blockchain. 

\textbf{G5 – Induce participants in the blockchain network to make errors}

There are several companies, municipalities and individuals producing transportation information which is valuable to governments, researchers and people. For example, one of the responsabilities of the government is to collect data in order to model, manage and improve transportation networks. Some malicious actors may inject the ecosystem with falsified transportation information via hacked traffic detectors, tolls, parking and smartcards. The participants of the blockchain network would end up making sub-optimal decision. The attackers may be interested in damaging the reputation of some company, or they may sow distrust in the government or any entity of the network. Attackers could include actors A1, A3, A4 and A5 identified in the previous section. Thus, inducing participants to make errors not only would they be achieving their goal, but they would also be evading the responsibilities of their actions by making their interference less detectable.

\begin{table*}[h]
	\centering
	\caption{Impact on the victims by attack goal - Monetary (M), Privacy (P), Integrity (I) and Trust (T). Impact scale ranges from 1 to 4, with 4 being the most severe.}
	\renewcommand{\arraystretch}{1}
	\begin{tabular}{|p{0.4\linewidth}|c|c|c|c|}
    \hline
\textbf{Attack goals} & \textbf{M} & \textbf{P}& \textbf{I} & \textbf{T}\\
    \hline
    G1-Gain knowledge about the data-market & 1 & 2 & -  & 1\\
    \hline
    G2-Access sensitive data on the nodes of the network & 2 & 3 & -  & 2\\
    \hline
    G3-Manipulate and modify blockchain information & 3 & 2 & 4 & 4\\
    \hline
    G4-Sabotage activities & 3 & - & 2 & 3\\
    \hline
    G5-Induce participants in the blockchain network to make errors & 2 & - & 3 & 3\\
    \hline
\end{tabular}
	\label{Impactbygoal}
\end{table*}

\subsection{Impact of attack goals }

Independently of the various actors goals, attacks will have an impact on the victim. The victim may be the individuals, the BSMD network or any of the active or passive node in the BSMD (data collectors, Companies, Universities and Government). In order to account for the various types of consequences that these attacks could have on them, we measure the impact according to four separate aspects: Monetary (M), Privacy (P), Integrity (I) and Trust (T). The impact scale ranges from 1 to 4, with 4 being the highest impact level (most severe). The impact levels are described in Table \ref{impactlevels}. The actor-based risk analysis is presented in Table \ref{Impactbygoal}. The explanation of the impact analysis by attack goal follows.

\textbf{G1} 	(M) Mobility data is very profitable, and competition between companies that seek to take advantage of the blockchain to use the data to improve their business is fierce. (P) While confidential, the information disclosed would not have severe consequences on the individuals. (T) The attack will have a minor impact on the trust in the BSMD network.

\textbf{G2} (P) While confidential, the information disclosed would not have severe consequences (except maybe in terms of insurability) and is likely to be otherwise available to actors through other more traditional forms of cyberattacks not related to the BSMD. (M) The disclosure of this information may be grounds for legal action against the agency, government bodies and concerned companies. (T) The attack will have a significant impact on the trust and will degrade confidence in the BSMD network.

\textbf{G3} (M) In the context of data markets, tampering of a service can produce losses of millions of dollars to companies. For example companies that take advantage of the blockchain to find customers. Also, government and transport agencies activities to improve mobility will suffer severe monetary losses. (P) Modifying blockchain information will have a significant impact on the privacy of any of the nodes in BSMD (Individuals, Companies, Universities and Government). (I) A tampered node's blockchain account will have catastrophic impact on the integrity of the mobility data, transactions and integrity of the user. (T) Similarly, the attack will have catastrophic impact on the trust of the BSMD network because the nodes will have no belief in the reliability or truth of the transactions in the ledger. 

\textbf{G4} (M) Malicious activities will induce severe monetary loss on the victim of the attack. For example, network hijacking, criminal smart contracts and ransomwares will exploit the victim in exchange of money. Also, sabotage activities will induce the vendor to not get rewards for its service. (I) The attacks will have a significant impact on the integrity of the mobility data, transactions and integrity of the users. (T) Finally, disruption of the network will have severe impact on the trust of the BSMD.

\textbf{G5} (M) Injecting the ecosystem with falsified transportation information will induce significant monetary loss on the victim of the attack. (I) The attack will have a severe impact on the integrity of the mobility data, transactions and integrity of the users. (T) Also, if participants relying on the mobility data acquired from the blockchain make errors in the modeling, management and improvement of transportation networks, it will degrade confidence and have a severe impact on the trust of the BSMD.

\subsection{Discussion}

This analysis responds to the needs of several groups such as companies, regulators, manufacturers, transportation agencies and even individuals. Each will be able to identify the riskiest threat, the one to treat with priority. From the results in Table \ref{Impactbygoal}, depending on the ranking obtained, a specific given risk management strategy can be applied. There are many strategies for managing the risk, namely: refuse, accept, transfer or reduce the risk. 

The most drastic one is refusing the risk. It is applied when the latter is unacceptable because of the catastrophic consequences it may have on the victims. The victims may be the individuals, universities, companies, transport agencies or any of the government nodes of the blockchain network. In this case, vulnerabilities in the blockchain system that are exploited by attackers should be removed from the system because of the security threat they pose. The strategy of accepting the risk is applied when the risk is either negligible or acceptable. In this case, the benefits that the system brings are greater than the potential risk. Transferring the risk strategy consists in giving the risk management responsibility to a third party, i.e. insurance companies. Finally, the risk mitigation strategy consists on reducing the risk as much as possible. This can be done through different means, such as the deployment of preventive mechanisms, security updates of the systems or stricter regulations.

In terms of monetary, privacy, integrity and trust, we note that G1 does not represent a potential risk that needs to be managed. However, G3 contains many threats that represent a risk that is either unacceptable or undesirable. 

From another angle, in terms of monetary impact, attack goals G2, G3, G4 and G5 represent a risk in terms of economic losses. For example, network hijacking, criminal smart contracts and ransomwares will exploit the victim in exchange of money. Also, sabotage activities will induce the vendor to not get rewards for its service. This monetary impact may concern certain groups  or organizations more than another. Thus, they would want to manage the risk by implementing the appropriate security defense mechanisms. For example, it is essential to ensure that cryptographic keys are stored or maintained properly so that the attacker aiming at accessing sensitive data on the nodes of the network do not exploit improper key protection mechanisms to attain the attack goal.

On the other hand, in terms of privacy impact, the results reveal that G2 is the riskiest attack goal. For example, attacks conducted by A1 may represent an undesirable risk that needs to be managed, i.e. cybercriminals that use compromised computer systems to commit identity theft. The compromise could lead to fraudulent transactions. Vulnerabilities should be eliminated by implementing efficient privacy preservation techniques and adopting more robust VPN solutions. 

In terms of integrity impact, particularly when it comes to the manipulation and modification of blockchain information, we notice from the results that this attack goal represents a major risk in terms of integrity. Also, this same attack goal has a catastrophic impact on the trust of the blockchain system because the nodes will have no belief in the reliability or truth of the transactions in the ledger. Among other solutions, threats should be managed by setting proper identity management (Membership Service Providers).

Finally, when it comes to inducing participants in the blockchain network to make errors, we notice from the results, that this attack goal represents a major risk in terms of integrity and trust in the blockchain ecosystem. To manage the risk, solutions should be implemented specifically regarding unauthenticated data feeds producing wrong data. 

\section{Conclusion}

We have proposed an actor-based risk assessment for the multi-layered Blockchain framework for Smart Mobility Data-markets (BSMD), a public-permissioned blockchain. Traditionally, the aim of the risk assessment is to evaluate the risk in order to identify the riskiest threat. The aim of our work is more extensive in that we propose a risk  analysis enabling the quantification of the risk associated not only to the blockchain technology, but also to its ecosystem. Moreover, we conducted the risk assessment methodology on a realistic blockchain for smart mobility data-markets with analysis of the attacks with regards to their impact at four scales (economy, privacy, integrity and trust). We quantify the risk by presenting ordinal values, which gives the decision makers a clear ranking in terms of prioritization. 

This work is the first step of a systematic risk assessment framework. We consider that although there are vulnerabilities in a system, it is their probability of exploitation and the impact that this exploitation has that determines whether the vulnerability represents a significant risk or not. In future work, we will extend the analysis and propose a scenario-based risk assessment followed by a combined risk assessment. The scenario-based analysis determines the probability of occurrence of each threat. The combined analysis aims at determining which attack outcomes have the highest risk according  to  their  impact  on  the  victims. We hope to uncover specific blockchain technology security vulnerabilities in the transportation ecosystem by exposing new attack vectors. The systematic risk analysis can then be used to develop possible countermeasures against cybersecurity vulnerabilities in the smart mobility implementations of the blockchain technology.

\bibliographystyle{ACM-Reference-Format}
\bibliography{Actor-basedanalysis}

\end{document}